\documentstyle[preprint,revtex]{aps}

\begin{document}

\begin{title}
Relativistic three-particle dynamical equations: \\ I. Theoretical development
\end{title}
\author{Sadhan K. Adhikari}
\begin{instit}
Instituto de F\'\i sica Te\'orica, Universidade Estadual Paulista \\
01405 S\~{a}o Paulo, S\~{a}o Paulo, Brasil
\end{instit}
 \moreauthors{T. Frederico}
\begin{instit}
Instituto de Estudos Avan\c cados, Centro T\'ecnico Aeroespacial \\
12231 S\~ao Jos\'e dos Campos,  S\~{a}o Paulo, Brasil
\end{instit}
\moreauthors{Lauro Tomio}
\begin{instit}
Instituto de F\'\i sica Te\'orica, Universidade Estadual Paulista \\
01405 S\~{a}o Paulo, S\~{a}o Paulo, Brasil
\end{instit}

\begin{abstract}
Starting from the two-particle Bethe-Salpeter equation in the ladder
approximation  and integrating over the time component of momentum, we rederive
three dimensional scattering integral equations satisfying constraints of
relativistic unitarity and covariance, first
derived by Weinberg and by Blankenbecler and Sugar. These two-particle
equations  are shown to be
related by a transformation of variables. Hence we show how to perform and
relate identical dynamical calculation using these two equations.
Similarly,
starting from the Bethe-Salpeter-Faddeev equation for the three-particle
system and integrating over the time component of momentum, we derive
several three dimensional
three-particle scattering equations satisfying constraints of
relativistic unitarity and covariance. We relate two of these  three-particle
equations by a transformation of variables as in the two-particle case.
The three-particle equations we derive are very practical and suitable for
performing relativistic scattering calculations.

\end{abstract}


\newpage

\vskip 5cm

\section{INTRODUCTION}

The physical basis for going beyond non-relativistic potential scattering is
secure for atomic processes where most phenomena are accurately described
within quantum electrodynamics (QED).\cite{Ad} In contrast to quantum
chromodynamics (QCD) in the hadronic and nuclear cases, QED is a weakly
coupled, non-confining field theory. It is relatively well understood how
to calculate physical quantities using the QED. For example, two-charged
particle scattering can be described by the covariant relativistic
Bethe-Salpeter (BS) equation,\cite{Bethe} which has the following generic
structure
\begin{equation}
 T= V +VG_0T,
 \label{BS}
 \end{equation}
 where the meaning and structure of the potential, $V$, and the Green function,
 $G_0$, are quite different than for simple potential scattering except the
 fact that $V$ has only potential cuts, and $G_0$ contributes to the unitarity
cuts of the $t$ matrix, $T$. The initial, final, and intermediate states of Eq.
(\ref{BS}) are defined in terms of four-momenta, and virtual processes are now
depicted as occurring off the mass-shell rather than off the energy-shell.

 It is still unclear on how to progress from QCD to practical collision
 integral equations for hadronic and nuclear processes. Therefore, most
 efforts to this end have employed a mixture of some type of meson-baryon
 field theory with phenomenology in order to obtain equations of the form
 (\ref{BS}), that presumably have a wider range of validity than the
 non-relativistic equations of the Lippmann-Schwinger (LS) type.\cite{Lipp}
  In this picture,  if Eq. (\ref{BS}) suffices to define $T$ uniquely,
then $V$ is necessarily an approximation.  There has been a number of
detailed calculations for  hadron-hadron collisions in the
 so-called ladder approximation of Eq. (\ref{BS}), where $V$ is approximated by
 the lowest order Feynman graphs.  Also, many approximation schemes have
 originated from the BS equation (\ref{BS}).\cite{RT} The full BS equation for
the few-hadron system correctly incorporates the fundamental constraints on
the  $t$ matrix, such as unitarity, particle creation etc. The BS equation,
 however, does not treat crossing symmetry exactly. Moreover,
 in the ladder approximation the BS equation does not satisfy unitarity
 at all energies, as they always contain some multi-particle contributions but
 not all. In our treatment, however, we shall be limited to a consideration of
a truncated Fock space with two or three particles and the ladder approximation
 to the BS equation satisfies unitarity below three- and four-particle
 thresholds.

The two-particle BS equation is a formal relation between vacuum
expectation values of time ordered products of covariant fields. None of
the covariant amplitudes in this equation is known, so one needs models for
them. One can make various truncations that reduce the BS equation to
a three-dimensional form. This reduction could be done in any choice of
variables.

 The most practical set of equations derived from the BS equation for the
 two- and the three-hadron systems are the instant-form
(constant-time) dynamical equations.
 \cite{BlS,AAY,Br} The use of instant-form variables
leads to dynamical equations quite similar
to the usual non-relativistic equations generated by Schr\"odinger dynamics
both for the two- and the three-particle systems and has been used virtually
in all numerical calculations of few-hadron scattering at medium energies.
\cite{Ad,BlS,AAY,Br,Rev,Tjon,Wall}
 The essence of the approaches leading to such equations is the replacement
 of the full four-dimensional Green function $G_0$ in the BS equation by an
 approximate three-dimensional one, consistent with conditions of two- and
 three-particle unitarity. This procedure reduces the full four-dimensional
 BS equations to three-dimensional LS-like equations\cite{Lipp} while
 preserving essential features like relativistic covariance, few-particle
 unitarity, and correct non-relativistic limit. Such dynamical equations
were first suggested by Blankenbecler and Sugar \cite{BlS} for the two-
particle system and later studied and extended to the three-particle system
by Aaron et al. \cite{AAY} However, these approximate
 equations develop  certain defects, not possessed by the original
 BS equation, such as, incorrect treatment of left-hand
 cuts.  The potential in the resulting three-dimensional equation is
 provided phenomenologically in order to fit experimental results. Practically
 all recent relativistic few-hadron scattering calculations have employed
 these approximate three-dimensional dynamical scattering equations in the
 instant-form.\cite{Tjon,Afnan,Garc}

The above three-dimensional reduction of the full four-dimensional BS equation
could be implemented in other variables also, apart from the normal
Schr\"odinger variables. Another set of variables is the
light-front variables, which we define in Sec. II. Weinberg\cite{Wein} was
the first to write relativistically covariant dynamical integral equations
for the two-particle system using light-front variables.
 The few-particle light-front dynamical equations are usually not derived from
 the BS equation as an approximation but obtained in an ad hoc fashion.
 \cite{Fuda,Bakk} The resulting equations, like the instant-form equations,
 are three-dimensional with a phenomenological potential.

The angular momentum decomposition in light-front variables, even for
potential scattering with spherically symmetric potential, is non-trivial and
this fact has severely limited the applications of the light-front dynamics
in actual physical problems of scattering.

In this paper
  we provide a general procedure for reducing four-dimensional
relativistic scattering integral equations, satisfying unitarity, to
three-dimensional forms. The present procedure consists of integrating over
the time component of the momentum variable in the intermediate state. In
the  case of the instant-form momentum variables this implies an integration
over the variable, $p_0$, whereas in the case of the light-front variables
the integration is to be performed over the equivalent variable, $p_-$.
While performing this integration
one assumes that the integrand - essentially, the potential and the $t$
matrix - is independent of the time component of momentum. There are
no other approximations
involved. Consequently, as the original four-dimensional
equations satisfy unitarity and relativity, so do the reduced ones.

Our main objective in this paper is to derive practical three-dimensional
three-particle scattering equations. However,
using the present procedure, as an excercise, first we derive
 dynamical equations for the two-particle
system starting from the BS equation.
For the two-particle system we recover the equations first derived by
Weinberg\cite{Wein} and by Blankenbecler and Sugar\cite{BlS}.
 We show that these two types of equations are related by a transformation
 of variables and hence are equivalent to each other. This also guarantees
 that the two-particle Weinberg equation yield unitary and rotationally
invariant result. This provides us with a recipe of performing dynamical
two-hadron scattering calculations using the Weinberg equation for a
specific angular momentum via the equivalent Blankenbecler-Sugar equation
where the angular momentum decomposition is under control.  In the
instant-form one has the usual one-dimensional partial-wave scattering
equation to solve, while in the light-front-form one has the three-dimensional
equation to solve. We show that both approaches lead to the same result. Our
finding holds good for any partial-wave potential, local or non-local.
We have worked out the algebra  for an S-wave separable potential.
We emphasize that we have not resolved the fundamental difficulties with
the angular momentum decomposition in light-front variables, but have found
a way to bypass them in solving the Weinberg equation. With the present
recipe the solution of the Weinberg and BlS equations are identical to each
other.

Next we employ the present procedure to the study of the
three-particle system employing two-particle separable potentials with an
objective to derive new three-particle scattering equations satisfying
constraints of relativistic unitarity and covariance. The separable form of
two-particle interaction simulates a two-particle bound state or an isobar.
In the presence of two-particle separable potentials
the three-particle equations are very similar to the usual two-particle
equations with non-local potentials. Hence the present derivation of the
 dynamical equations for the two-particle system
from the BS equation is readily applicable to the three-particle system.
We derive several  instant-form
and one light-front scattering equations for the
three-particle system. Three-dimensional three-particle scattering equations
have also been proposed in Refs. \cite{BlS,AAY,Tjon,Fuda,Bakk,Fred}
One of the present  three-particle
instant-form equations we derive is a consequence of using a three-particle
relativistic propagator derived by
 Ahmadzadeh and Tjon\cite{Tjon} and is shown
to be related to the present light-front equation by a transformation of
variables. The present instant-form equations are distinct from the
dynamical equations derived
by Aaron, Amado, and Young\cite{AAY}.

The instant-form equations we derive are very suitable for performing
numerical calculation as we demonstrate for the three-nucleon system in the
following paper.\cite{our}

The plan of the paper is as follows. In Sec. II we present the essential
kinematics for the two-particle system, derive the instant-form and
light-front equations for the
two-particle system starting from the BS equation,
and show how and under what conditions one can pass from one form to the
other. This allows us to perform a numerical calculation using the light-front
equation via the equivalent equation in the instant form. We illustrate these
ideas using a separable two-particle potential. In Sec. III we generalize the
treatment of Sec. II to the case of three-particles with separable
two-particle  potentials. In particular we derive new instant-form and light
front equations for the three-particle system, each satisfying two- and
three-particle   unitarity.  Finally, in Sec. IV we present some concluding
remarks.

\section{The two-particle problem}

Our principal interest in the present study is to develop practical
three-dimensional scattering integral equations for the three-particle system,
which can be used easily in numerical calculations. We present a numerical
application using these equations in the following paper.\cite{our} However,
in this section we find it convenient to illustrate our approach for the
two-particle system, which is well-understood, before taking up the
challanging task of deriving three-dimensional three-particle equations in
the following section.

\subsection{Kinematics}

The instant-form dynamics is treated in terms of the usual four vectors
$(x_0,\vec x)$ and $(p_0,\vec p)$ in the configuration and momentum spaces,
respectively. Here $\vec x \equiv (x_1,x_2,x_3)$,
$\vec p \equiv (p_1,p_2,p_3)$, where $x$ and $p$ denote position and momentum
components; $x_0$ is time and $p_0$ represents energy.  The invariant length
square of the four vector $x$ is given by $x^2 = (x_0^2 -\vec x ^2)$.
We adopt units $\hbar =c=1$. The light-front dynamics is treated
in terms of equivalent momentum variables $(p_+,p_-,\vec p_\perp)$, with
$\vec p_\perp \equiv (p_1,p_2)$, $p_+ =p_0+p_3$, and $p_-=p_0-p_3$. The on the
mass shell condition for a particle of mass $m$ in the two systems are given by
\begin{equation}
 p_0^2 =  {\vec p}^2 +m^2,
\label{1}
\end{equation}
and
\begin{equation}
      p_+p_- = {\vec p_\perp}^2 +m^2,
\label{2}
\end{equation}
in the instant-form and light-front formalisms, respectively.
One also has the following useful identity
\begin{equation}
p^2 -m^2 = p_+(p_- -\frac {{\vec p_\perp}^2 +m^2}{p_+}).
\label{XXX}
\end{equation}

For a system of two particles the invariant energy
square, $s$, is given by
\begin{eqnarray}
s & = & P_+P_-  - \vec P_ \perp ^2 \\
  & = & \sum_i \frac {m_i^2+\vec p_{i\perp}^2}{{p_i}_+} P_+ -\vec P_\perp ^2\\
  & = & \sum_i \frac {m_i^2+\vec p_{i\perp}^2}{x _i} -\vec P_\perp ^2,
  \label{3}
  \end{eqnarray}
where $x _i = {p_i}_+/P_+$ is the momentum fraction of particle $i$,
$p_i$ and $m_i$
denote the momentum and mass of individual particles, the index $i$ labels
the particles,  $i=1,2$. The variable $P$ refers to the total four momentum of
 the system of two particles, $P=p_1+p_2$ and its components are defined in
 close analogy to those of the individual particle.
  It is convenient to define the relative momentum variable
  \begin{equation}
  \vec k _\perp =x_1\vec p_{2\perp} -x_2\vec p_{1\perp},
\label{4}
\end{equation}
where $x_2=(1-x_1)$. In terms of this variable the invariant energy
square is given by
\begin{equation}
s=\frac{m_2^2}{x_2} +\frac{m_1^2}{x_1}+\frac{\vec k_\perp^2}
{x_1 x_2}.
\label{A}
\end{equation}
This equation has the same form in all frames and this is one of the
advantages of working with the light-front variables.

In the case of the instant-form variables it is convenient to work in the
c.m. system, so that $\vec P =0$ and $P_0=\sqrt s$,
and define the relative four
momentum $p$ by $2p=p_1-p_2$ in close analogy with non-relativistic
kinematics. The individual particle momenta are hence given by $p_1=
P/2+p$ and $p_2=P/2-p$. We shall always be working  in the c.m.
system in both instant-form and light-front formalisms.

For instant-form and light-front variables the invariant energy square in the
c.m. frame is given by
\begin{equation}
s = [(\vec p^2+m_1^2)^{1/2} + (\vec p^2 +m_2^2)^{1/2}]^2,
\label{A1} \end{equation}
and
\begin{equation}
s=\frac {m_1^2+\vec p_{\perp}^2} {x_1}+\frac {m_2^2+\vec p_{\perp}^2} {x_2}.
\label{A2} \end{equation}

With  this discussion of kinematics for two noninteracting particles we
shall now present a discussion of the dynamics of two interacting particles
in the next subsection.

\subsection{Instant-form equations}

As we base the unified derivation on the BS equation it is convenient to write
the approximate form of the BS equation for the two-hadron system in the form
\begin{equation}
t(q,k,s) = V(q,k,s)+\frac {i}{(2\pi)^4}\int d^4p
\frac {V(q,p,s)t(p,k,s)} {(p_1^2-m_1^2+i0)(p_2^2-m_2^2+i0)}.
\label{BS2}
\end{equation}
The usual derivation of the instant-form equations approximates the
two-particle Green function
\begin{equation}
G_0(p,s)\equiv (p_1^2-m_1^2+i0)^{-1}(p_2^2-m_2^2+i0)^{-1}
\label{B}
\end{equation}
by a three-dimensional one  satisfying unitarity, sets all the
particles (even in the intermediate state) on the mass shell,
 and assumes further that the
potential and the $t$ matrix are independent of the time component of
momentum variables involved.

If we assume in Eq. (\ref{BS2}) that the potential
and the $t$ matrix are independent of the time component of momenta
variables, the $dp_0$ integration can
 be readily performed and one arrives at
an instant-form equation which is identical to the three dimensional equation
of Blankenbecler and Sugar and of Aaron, Amado, and Young.\cite{BlS,AAY}
No further approximations on the Green function is needed for the purpose.
As the original equation (\ref{BS2}) satisfy unitarity and relativity,
so do the reduced ones.

The same procedure could be carried through in the light-front variables, too.
This procedure allows us to establish an equivalence between the light
front and instant-form dynamical equations not only for the two-particle
system but also for the three-particle system.

For the two particles in the intermediate
state in Eq. (\ref{BS2}) we have
\begin{equation}
p_1^2 -m_1^2 = (p_0+\frac{\sqrt {s}}{2})^2 -\omega _1 ^2,
\label{C}
\end{equation}
and
\begin{equation}
p_2^2 -m_2^2 = (p_0-\frac{\sqrt {s}}{2})^2 -\omega _2 ^2,
\label{D}
\end{equation}
where $\omega _i^2 = \vec p^2+m_i^2, i=1,2,$ are the squares of energy of
the two particles.
Hence the last term in the BS equation (\ref{BS2}) is written as
\begin{eqnarray}
VG_0t & = &  \frac{i}{(2\pi)^4} \int d \vec p \int _{-\infty} ^{\infty} dp_0
\frac   {V(\vec q,\vec p,s)t(\vec p,\vec k,s)} {(p_0+\sqrt{s}/2+\omega _1 -i0)
(p_0+\sqrt{s}/2-\omega _1 +i0)} \nonumber \\
   & \times & \frac {1} {(p_0-\sqrt{s}/2+\omega _2 -i0)
(p_0-\sqrt{s}/2-\omega _2 +i0)},
\label{E}
\end{eqnarray}
where we have assumed that the potential, $V$, and the $t$ matrix are
independent of the time component of the four momentum.  The evaluation
of the $dp_0$ integral in Eq. (\ref{E}) is a
straightforward exercise in the analysis of complex variables using the
Cauchy theorem. If the contour of $p_0$ integration  is taken along the
real axis from $-\infty$ to $\infty$ and closed in
the counterclockwise sense along a semicircle in the upper half of the complex
$p_0$ plane at infinity this  integral is readily evaluated to yield
\begin{eqnarray}
VG_0t & = &-\frac{1}{(2\pi)^3}\int d\vec p\biggr[\frac{1}{(-2\omega_1)
(-\sqrt{s}-\omega_1+\omega_2)(-\sqrt{s}-\omega_1-\omega_2-i0)} \nonumber \\
 & + & \frac{1}{(-2\omega_2)(\sqrt{s}+\omega_1-
\omega_2)(\sqrt{s}-\omega_1-\omega_2+i0)}\biggr] V(\vec q,\vec p,s)
t(\vec p,\vec k,s).
\label{F}
\end{eqnarray}
This last result can easily be simplified and consequently the BS equation
(\ref{BS2}) reduces to
\begin{equation}
t(\vec q,\vec k,s) = V(\vec q,\vec k,s) +\frac{1}{(2\pi)^3} \int d\vec p
V(\vec q,\vec p,s) \biggr[ \frac{\omega_1+\omega_2}{2\omega_1 \omega_2[s-
(\omega_1+\omega_2)^2+i0]}\biggr] t(\vec p,\vec k,s). \label{G}
\end{equation}
This equation was first derived explicitly by Aaron, Amado, and Young\cite{AAY}
for different mass particles using the procedure of Blankenbecler and Sugar.
\cite{BlS} Equation (\ref{G}) is three dimensional, covariant and satisfies
constraints of unitarity. After a partial-wave projection the solution of
this equation is easily related to scattering phase shifts.

The energy denominator of Eq. (\ref{G}) has two poles. One of them corresponds
to the condition $\sqrt s = (\omega _1 +\omega _2) $, and represents the
propagation of two particles in the intermediate state. The other corresponds
to the condition $\sqrt s =- (\omega _1 +\omega _2) $, and represents the
propagation of two antiparticles. The first pole is responsible for
maintaining unitarity in the two particle sector. Usually, all three
dimensional reductions of relativistic scattering equations involve propagation
of antiparticle(s) in the intermediate state. However, the residue of the
quantity in the square bracket in Eq. (\ref{G}) at the pole corresponding to
the particle propagation has to be the same in all formulations, which is the
condition for maintaining unitarity in the two-particle sector. Bakker et
al.\cite{Bakk}
have suggested to eliminate the propagation of antiparticles by evaluating
the quantity in the square bracket of Eq. (\ref{G}) at the pole corresponding
to propagation of particles. Consequently, Eq. (\ref{G}) reduces to
\begin{equation}
t(\vec q,\vec k,s) = V(\vec q,\vec k,s) +\frac{1}{(2\pi)^3} \int d\vec p
V(\vec q,\vec p,s) \biggr[ \frac{1}{4\omega_1 \omega_2[\sqrt s-
(\omega_1+\omega_2)+i0]}\biggr] t(\vec p,\vec k,s). \label{GGG}
\end{equation}
Equation (\ref{GGG}) further simplifies to Eq. (2.36) of Ref.\cite{Bakk}
for equal
mass particles, when the quantity in the square bracket of this equation is
evaluated at  the pole corresponding to particle propagation.

\subsection{Light-front equations}

Now we shall carry on the reduction procedure of the last subsection using the
light-front variables. For the two particles
in the intermediate state in Eq. (\ref{BS2}) we now have
\begin{equation}
p_1^2 -m_1^2 +i0= (p_++\frac {\sqrt{s}}{2})\biggr[ (p_-+\frac {\sqrt{s}}{2})
    -\frac {\vec p_\perp^2+m_1^2-i0}{p_++\frac {\sqrt{s}}{2}}\biggr] ,
    \label{H}
    \end{equation}
    and
\begin{equation}
p_2^2 -m_2^2 +i0= (-p_++\frac {\sqrt{s}}{2})\biggr[ (-p_-+\frac {\sqrt{s}}{2})
    -\frac {\vec p_\perp^2+m_2^2-i0}{-p_++\frac {\sqrt{s}}{2}}\biggr] .
\label{I} \end{equation}
 If we use Eqs. (\ref{H}) and (\ref{I}), the BS equation
(\ref{BS2}) becomes
\begin{eqnarray}
t(q,k,s) & = & V(q,k,s)+  \frac {i}{2(2\pi)^4}\int d\vec p_\perp
\int _{-\infty } ^\infty dp_+  \int _{-\infty } ^\infty dp_-
\frac {V(q,p,s)t(p,k,s)} {(p_++\frac {\sqrt{s}}{2})[(p_-+\frac {\sqrt{s}}{2})
    -\frac {\vec p_\perp^2+m_1^2-i0}{p_++\frac {\sqrt{s}}{2}}]} \nonumber \\
& \times & \frac {1} {(-p_++\frac {\sqrt{s}}{2})[(-p_-+\frac {\sqrt{s}}{2})
    -\frac {\vec p_\perp^2+m_2^2-i0}{-p_++\frac {\sqrt{s}}{2}}]}.
  \label{J}
  \end{eqnarray}
In Eq. (\ref{J}) the factor of (1/2) before the integral is the Jacobian of
the transformation of integration variables from $p\equiv (p_0,\vec p)$ to
$p\equiv (p_-,p_+,\vec p_\perp)$. Also, a typical momentum four vector in
this equation, e.g., $q$, is considered to have components
$(q_-,q_+,\vec q_\perp)$. These components are to be contrasted with the
usual components $(q_0,\vec q)$. Of the light-front momentum variables, $q_-$
plays the role of the time  component of the momentum four vector,
and following the procedure of the last subsection we would like to perform
the $p_-$ integration in Eq. (\ref{J}).
In the context of the light-front variables this integration can be performed
if we take the $t$ matrix and the potential to be  independent of the
time component of momentum. We assume that this is the case. A similar
assumption was made while performing the $p_0$ integration in Eq. (\ref{E}).

We note that the $dp_-$ integral in Eq. (\ref{J}) can be transformed into a
contour integral of the form
\begin{equation}
\oint  dz \frac {1} {(z-z_1)(z-z_2)},
\label{K}
\end{equation}
where the contour goes along the real $z$ axis from  $x=-\infty$ to $x=\infty$
and is closed in the counterclockwise sense along a semicircle in the
upper-half complex $z$ plane at infinity, with $x$ being the real part of $z$.
The integral (\ref{K}) contributes when the contour includes only one of
the poles of the integrand.  Assuming that it includes the pole $z=z_1$, the
result of the integral (\ref{K}) is $2\pi i/(z_1-z_2)$.

The condition of including one of the poles of $p_-$ of Eq. (\ref{J}) inside
the contour sets for positive $\sqrt s$ the following limit on $p_+$:
$\sqrt s/2 > p_+ >-\sqrt s/2$. The integral over $p_-$ of Eq. (\ref{J}) is
now readily evaluated to yield
\begin{eqnarray}
VG_0t & = & \frac{1}{2} \frac {i}{(2 \pi)^4}\int d \vec p_\perp
\int_{-\sqrt s/2}
^{\sqrt s/2} dp_+\frac {-2\pi i V(q_+,\vec q_\perp;p_+,\vec p_\perp;s)
t(p_+,\vec p_\perp;k_+,\vec k_\perp;s)} {(p_++\sqrt s/2)(-p_++\sqrt s/2)}
\nonumber  \\
& \times &\frac {1} {\sqrt s -\frac {\vec p_\perp^2+m_2^2}{-p_++\frac
{\sqrt{s}}{2}}
-\frac {\vec p_\perp^2+m_1^2}{p_++\frac {\sqrt{s}}{2}}+i0}.
\label{L} \end{eqnarray}
It is convenient to rewrite Eq. (\ref{L}) in terms of the momentum fraction
$x_p$
defined by
\begin{equation}
x_p = \frac {p_+ +\sqrt s/2}{\sqrt s},
\label{M} \end{equation}
with other momentum fractions $x_q$, $x_k$, etc. defined analogously. In terms
of these momentum fractions Eq. (\ref{L}) is easily simplified and the BS
equation (\ref{BS2}) is rewritten as
\begin{eqnarray}
t(x_q,\vec q_\perp;x_k,\vec k_\perp;s) & = & V(x_q,\vec q_\perp;x_k,\vec
k_\perp;s)
+ \frac {1} {(2\pi)^3} \int _0^1 \frac
{dx_p}{2x_p(1-x_p)}\int d\vec p_\perp  \nonumber \\
& \times & \frac {V(x_q,\vec q_\perp;x_p,\vec p_\perp;s)
t(x_p,\vec p_\perp;x_k,\vec k_\perp;s)}  {s-\frac {\vec p_\perp^2+m_2^2}
{1-x_p} -\frac{\vec p_\perp^2+m_1^2}{x_p}+i0}.
 \label{N} \end{eqnarray}
Equation  (\ref{N}), first derived by Weinberg,\cite{Wein} is the light-front
counterpart of the relativistic LS-type equation (\ref{G}) in the instant
form.  This equation is not rotationally invariant in light-front variables
and hence the  physical content  and interpretation of this equation and its
relation to Eq. (\ref{G}) is unclear.  Related to this is the treatment of
angular momentum in Eq. (\ref{N}). We shed light on  some of these problems
in the next subsection. Specifically, we show the equivalence between the
physical contents of  Eqs. (\ref{G}) and (\ref{N}). As an illustration we
consider an $S$ wave separable potential in order to clarify some of these
issues.

\subsection{Equivalence between light-front and instant-form equations}

We begin this subsection by demonstrating the equivalence between the above
two sets of equations (\ref{G}) and (\ref{N}).
In order to achieve our goal we would like to find a possible  transformation
of variable, $x_p$ to $p_3$, which relates these two equations without any
other approximation,
whatsoever. We used Eq. (\ref{M}) to derive (\ref{N}). In Eq. (\ref{N}),
however, $x_p$ is just a variable of integration. We recall that we have two
on the mass shell particles in the intermediate state of four momentum given by
$(\omega _1,\vec p)$ and $(\omega _2,-\vec p)$ in the c. m. frame with
$\omega_i=(\vec p^2+m_i^2)^{1/2}$, $i$=1,2. Physically, $x_p$ represents the
momentum fraction of the first particle and should be given by
\begin{equation}
x_p = \frac {\omega_1 + p_3} {\omega_1+\omega_2}.
\label{O} \end{equation}
We shall see that transformation
(\ref{O}) will take us from the light-front equation (\ref{N}) to the instant
form equation (\ref{G}) with some plausible conditions on the potentials
and the $t$ matrices.

The Jacobian of the transformation (\ref{O}) is given by
\begin{equation}
\frac{dx_p}{dp_3}  =  x_p(1-x_p) \frac{\omega_1+\omega_2}{\omega_1 \omega_2}.
\label{P} \end{equation}
We also note that for $x_p$ given by Eq. (\ref{O}) one has the following
identity
\begin{equation}
\frac {\vec p_\perp^2+m_2^2}{1-x_p}
 +\frac{\vec p_\perp^2+m_1^2}{x_p} = (\omega _1 +\omega _2)^2.
\label{PPP} \end{equation}
If we use Eqs. (\ref{P}) and  (\ref{PPP}) in  the light-front equation
(\ref{N}), we arrive at the instant-form equation (\ref{G}), provided
that the potential and the $t$ matrix transform according to:
\begin{equation}
V(x_q,\vec q_\perp;x_p,\vec p_\perp;s) \to V(\vec q, \vec p;s),
\label{Q} \end{equation}
\begin{equation}
t(x_q,\vec q_\perp;x_p,\vec p_\perp;s) \to t(\vec q, \vec p;s).
\label{R} \end{equation}
This establishes the desired equivalence. The instant-form equation
(\ref{G}) is manifestly rotationally invariant. The present equivalence
guarantees that the light-front equation (\ref{N}), though not rotationally
invariant in light-front variables, becomes so once transformed to instant
form. In other words, the light-front equation (\ref{N}) should yield
rotationally invariant result
in normal instant-form variables. We shall explicitly demonstrate
this in the next subsection for a separable potential.

We emphasize that the physical contents of the two equations (\ref{G}) and
(\ref{N}) are distinct. However, the potentials in these two equations are
not derived from any consistent theory. They are phenomenological
inputs, considered as ad hoc approximations to the potential of the BS
equation, designed to generate experimental observables via the two types
of equations. By construction, both yield relativistically covariant result
with correct non-relativistic limit. This result is rotationally invariant
and unitary, as we shall see below,
 when expressed in physical instant-form variables.

The light-front equation is not manifestly spherically symmetric, and this
leads to complications in angular momentum projection and in the
interpretation of solutions of this equation. The present equivalence
leads to a recipe to bypass
both these problems. One possibility is to first transform
the light-front potential to the instant form via (\ref{Q}) and then solve the
equivalent instant-form equation. The other possibility is to solve the
three dimensional light-front equation without attempting a partial wave
projection for each normal physical partial wave concerned. We shall see
in the following that the two possibilities  lead to the same result.

The main non-trivial question is how to effect the transformation (\ref{Q}).
This is achieved through the use of Eq. (\ref{PPP}), which
 gives the unique transformation $(x_p\to p_3)$ that relates the potentials
 in the two frames. However, for different mass particles this transformation
 is quite involved. This is why in the following we specialize to the case of
 equal mass particles, $m=m_1=m_2$, where Eq. (\ref{PPP}) reduces to
\begin{equation}
\vec p^2= \frac {\vec p_\perp^2+m^2}{4x_p(1-x_p)} -m^2.
\label{T} \end{equation}
Equation (\ref{T}) provides the unique transformation required to construct
the light-front potential from the instant-form potential in each physical
partial wave. The partial wave projection in Eq. (\ref{G}) for the equal mass
case can be readily carried through and one has the following partial wave
form
\begin{equation}
t_l(q,k,s)=V_l(q,k,s)+\frac {1}{4(2\pi)^3} \int _0 ^\infty \frac {4 \pi p^2 dp}
{(p^2+m^2)^{1/2}} V_l(q,p,s)  \frac{1}{(k^2-p^2+i0)} t_l(p,k,s),
\label{U} \end{equation}
where $l$ denotes the partial wave. In the partial wave equations, such as Eq.
(\ref{U}) and in the following, the momentum labels $p, q$ etc. denote the
modulus of the corresponding three vectors. In arriving at Eq.(\ref{U}) we
have set
the variable $k$ on the energy shell too, so that $s=4(m^2+k^2)$. Though there
are many ways of making the partial wave projection, we have written the
partial wave equation in such a way that the full phase space, $4\pi p^2 dp$,
appears in this equation. This has no relevance on our conclusions. However,
it makes our case more transparent, as Eq.(\ref{U}) is  the full
equation for a potential with just a $l$ wave component, and the full
equation has been demonstrated to be equivalent to the light-front equation.

It is interesting to note that in the non-relativistic limit the factor
$(p^2+m^2)^{1/2}$ in Eq. (\ref{U}) reduces to $m$ and this equation reduces
to the usual partial wave LS equation.

For the sake of completeness, we observe that in the case  of equal mass
particles the light-front equation  (\ref{N}) reduces to
\begin{eqnarray}
t(x_q,\vec q_\perp;x_k,\vec k_\perp;s) & = & V(x_q,\vec q_\perp;x_k,\vec
k_\perp;s)
+ \frac {1} {2(2\pi)^3} \int _0^1 \frac
{dx_p}{x_p(1-x_p)} \int d\vec p_\perp \nonumber \\
& \times & \frac {V(x_q,\vec q_\perp;x_p,\vec p_\perp;s)
t(x_p,\vec p_\perp;x_k,\vec k_\perp;s)}  {s -\frac {\vec p_\perp^2+m_2}
{x_p(1-x_p)}+i0}.
 \label{U1} \end{eqnarray}

Given a spherically symmetric potential with component
$V_l(q,p;s)$ in partial wave $l$, a
numerical calculation with the partial wave form (\ref{U}) can be performed
in a routine fashion. The transformation (\ref{T}) provides the unique
prescription of constructing the light-front (LF) potential, to be
employed in Eq. (\ref{U1}),  from the
instant-form (IF) potential for the physical partial wave $l$ via
\begin{eqnarray}
V_l^{(LF)}(x_q,\vec q_\perp;x_p,\vec p_\perp;s) & = & V_l^{(IF)}(q,p;s)
\nonumber \\
& = & V_l^{(IF)}\left(\left[\frac {\vec q_\perp^2+m^2}
{4x_q(1-x_q)} -m^2\right]^{1/2},
\left[\frac {\vec p_\perp^2+m^2}{4x_p(1-x_p)}
-m^2\right]^{1/2};s\right),\nonumber \\
&     &
\label{V} \end{eqnarray}
where we have used an equation of the type (\ref{T}) for each of the
momentum variables. Also we have used the indices LF  and IF on potentials
in order to be more specific. The present demonstration of equivalence
of the two sets of equations together with prescription (\ref{V}) suggest
two equivalent calculational schemes in the two approaches which would lead to
identical results.  The first possibility is to solve the one dimensional
instant-form equation (\ref{U}) with potential $V_l^{(IF)}(q,p;s)$, and
the second
possibility is to solve the three dimensional light-front equation (\ref{U1})
with the potential $V_l^{(LF)}(x_q,\vec q_\perp;x_p,\vec p_\perp;s)$ given
explicitly by the right hand side of Eq. (\ref{V}). Once the functional
form of the instant-form potential $V_l^{(IF)}(q,p;s)$ is known the
equivalent light-front potential could be readily found out via (\ref{V}).

We note again that we have not resolved the difficulties
with the angular momentum projection of the light-front equations.  The
light-front equation still continues to be three dimensional. We have provided
a recipe which allows to perform scattering calculations employing partial-wave
instant-form potential via BlS and Weinberg equations. Provided that the
potentials in these two equations are related by Eq. (\ref{V}), both the
equations lead to the same result.
Our conclusions are quite general. But as they are quite subtle, a
simple and specific example would help make our issue. And we do this in the
next  subsection  considering the analytic separable $S$ wave potential model.

\subsection{$S$ wave potential model: an illustration}

Separable potentials have been
successfully used in diverse situations.  We consider
the following $S$ wave energy independent separable two-hadron potential in
the momentum
space to be used in the partial wave instant-form equation (\ref{U})
\begin{equation}
V_0(|\vec M_q|,|\vec M_p|,s) \equiv V(|\vec M_q|,|\vec M_p|)
= \lambda g(|\vec M_q|^2) g(|\vec M_p|^2
)
\label{W} \end{equation}
where $\vec M_p$ is a special vector and is a function of momenta  $p_1$ and
$p_2$ of the two particles. This is in fact the relative
momenta of the two particles in the c. m. frame. In order to maintain Lorentz
covariance of the equations, the best thing is to introduce a Lorentz
invariant in place of $|\vec M_p|$. The Lorentz invariant, which reduces to
the relative momentum square in the c. m. frame is $[(p_1-p_2)/2]^2$. For equal
mass particles  in the c. m. frame $|\vec M_p |^2 = \vec p^2$.
Then Eq. (\ref{W}) can be rewritten in the c. m. frame as
\begin{equation}
V(|\vec M_q|,|\vec M_p|)
= \lambda g(\vec q ^2) g(\vec p ^2).
\label{WW} \end{equation}
We assume the following simple form factors  for the potential
\begin{equation}
   g(\vec p ^2) = \frac {1}{(\vec p^2+\beta ^2)^{1/2}}.
   \label{X} \end{equation}
 With this potential the $t$ matrix of equation    (\ref{U}) is readily
 given by
 \begin{equation}
 t_0(q,p,\vec k^2) = \tau (\vec k^2) g(\vec q ^2) g(\vec p ^2),
\label{Y} \end{equation}
where
\begin{equation}
\tau ^{-1} (\vec k^2) = \frac {1}{\lambda } -\frac {1}{8\pi ^2} \int _0 ^\infty
{p'}^2 dp' \frac {g^2({\vec {p'}}^2)}{({\vec {p'}}^2+m^2)^{1/2}(\vec k^2-
{\vec {p'}}^2-+i0)}.
\label{Z} \end{equation}
In Eqs. (\ref{Y}) and (\ref{Z}) $\vec k$ is on the energy shell relative
momentum in the c. m. frame: $ s = 4(\vec k ^2 +m^2)$.

Such a $t$ matrix is very plausible when the two-particle interaction
proceeds through coupling to a definite isobar in a particular partial wave
only. To make the algebra simple we will assume that all the
particles considered are spin and isospinless and the isobar occurs in
relative $S$ wave only.  If desired, these restrictions could be removed in
principle without too much trouble. The function  $\tau ^{-1} (\vec k^2)$
should have the form $(s-m_I^2)$
where $m_I$ is the dressed mass of the isobar. We note that the function
$\tau ^{-1} (\vec k^2)$ has a zero at the isobar mass, carries the scattering
phase, and has the unitarity cut, whereas the function $g$ should yield the
left-hand cut.

With the form factor of Eq. (\ref{X}) the integral in Eq. (\ref{Z}) could be
evaluated to yield
\begin{eqnarray}
\tau ^{-1} (\vec k^2) & = & \frac {1}{\lambda }+\frac {1}{8\pi ^2(\beta^2+\vec
 k ^2)}\biggr[\frac{\beta}
{(m^2-\beta^2)^{1/2}} arctan\frac{(m^2-\beta^2)^{1/2}}{\beta} \nonumber \\
  & + &\frac{ik\pi}{2(m^2+\vec k ^2)^{1/2}} +\frac{k}{2(m^2+\vec k ^2)^{1/2}}
ln \frac{ (m^2+\vec k ^2)^{1/2} -k}{(m^2+\vec k ^2)^{1/2} +k}\biggr].
\label{1A} \end{eqnarray}
Equations (\ref{Y}) - (\ref{1A}) provide the usual solution to the problem
via the partial  wave instant-form equation (\ref{U}). In Eq. (\ref{1A})
the symbol $k$
denotes the modulus of the on shell relative three momentum $\vec k$ in the
c. m. frame.

The same problem could also be solved and the same solution obtained via
the full three dimensional light-front equation (\ref{N}). The potential to
be used in this equation, given by prescription (\ref{V}), is written
explicitly as:
\begin{equation}
V(x_q,\vec q_\perp;x_p,\vec p_\perp;s)=\lambda g(x_p,\vec p_\perp)
g(x_q,\vec q_\perp),
\label{1B} \end{equation}
where
\begin{equation}
g(x_p,\vec p_\perp)=\biggr[\frac {\vec p_\perp^2+m^2}{4x_p(1-x_p)}
-m^2+\beta ^2\biggr]^{-1/2}.
\label{1C} \end{equation}
This potential is separable and the solution of Eq (\ref{N}) could be found by
analytic means. The solution is given by
\begin{equation}
t(x_q,\vec q_\perp;x_p,\vec p_\perp;s)=\tau (\vec k^2)  g(x_p,\vec p_\perp)
g(x_q,\vec q_\perp),
\label{1D} \end{equation}
where
\begin{equation}
\tau ^{-1} (\vec k^2) =\frac {1}{\lambda } -\frac{1}{16\pi ^3}
\int _0^1 \frac
{dx_p}{x_p(1-x_p)} \int d\vec p_\perp \frac {g ^2(x_p,\vec p_\perp)}
{s -\frac {\vec p_\perp ^2+m^2}{x_p(1-x_p)}}.
\label{1E} \end{equation}
With the form factor given by Eq. (\ref{1C}) the integral in Eq. (\ref{1E})
could be analytically evaluated. We note that the integrand is independent of
the angles of the vector $\vec p_\perp$. If a transformation of variables to
$z = (\vec p_\perp ^2 +m^2)/[x_p(1-x_p)]$ is made and $x_p$ is assumed to be
constant, the integral over $p _\perp $
could be evaluated to yield
\begin{equation}
\tau ^{-1} (\vec k^2) =\frac {1}{\lambda } +\frac{1}{16\pi ^2}
\frac{4}{(4\beta ^2 -4 m^2
+s)} \int _0 ^1 dx ln \frac{m^2 +4(\beta ^2 -m^2)x(1-x)}{m^2 -sx(1-x)}.
\end{equation}
Using integrals [2.733] and [2.736] of Ref. \cite{Grad}
 the remaining integral could
be evaluated and after some lengthy but straightforward algebra one arrives
at the final rotationally invariant result (\ref{1A}). This also provides an
explicit demonstration
of rotational invariance of the light-front equation (\ref{U1}) in
instant-form variables. Though the light-front equation is not rotationally
invariant in light-front variables, it  yields rotationally invariant
result when expressed in instant-form variables.

With  condition (\ref{T}) the form factors (\ref{X}) and
(\ref{1C}) are identical. Hence, the two $t$ matrices given by  Eqs. (\ref{Y})
and (\ref{1D}) are also identical.
This demonstrates the equivalence between the
instant-form and the light-front equations for the present separable potential.
Of course, the equivalence is valid for any spherically symmetric partial wave
potential - local or non-local.  In the instant form one has to solve the
partial wave
one dimensional integral equation (\ref{N}), whereas in the light-front
formalism one has to solve the three dimensional equation (\ref{U1}). Both
procedures lead to the same result subject to condition
(\ref{T}).

\section{the three-particle problem}

We now use the procedure of the last section to obtain a set of relativistic
three-particle equations both in the instant-form and light-front formalisms.
Again for simplicity we shall consider the case of three identical
spin and isospinless bosons.
In stead of considering the general problem of three particle scattering
we shall consider the problem of elastic scattering of one particle from the
bound state of the other two. We also assume
 an isobar in the two-particle system.

\subsection{Instant-form equations}

Just as in the two-particle case we assume a form for the
four-dimensional relativistic Bethe-Salpeter-Faddeev three-particle equations
for three identical spin and isospinless bosons,
which we take as
\begin{equation}
T(q,k,s) = 2B(q,k,s) + \frac {2i}{(2\pi)^4}\int d^4 p B(q,p,s)
\tau (\sigma _p) T(p,k,s),
\label{1F}
\end{equation}
with
$\sigma _p \equiv (P-p) ^2 = (\sqrt s -\omega _{\vec p})^2 - \vec p ^2,$
where $P$ is the total four momentum of the system of three particles in the
c. m. frame and is given by $(\sqrt s, 0 , 0, 0)$. The factor 2 in the
Born and  the homogeneous terms of Eq. (\ref{1F}) is due to symmetrization
for identical bosons.
Equation (\ref{1F}) is a straightforward generalization of the non-relativistic
Amado model, which has been shown to be equivalent to the Faddeev equation for
separable two-particle potential. We note that Eq. (\ref{1F}) is not derived
from some fundamental theory, but its structure is
postulated from those of the two-particle BS equation
and  the non-relativistic Amado model.\cite{Amado}
In Eq. (\ref{1F}) $\tau$ has the
same meaning as in Eq. (\ref{Y}).  The internal dynamics of the two-particle
sub-system given by $\tau$ is supposed to be provided, so that Eq. (\ref{1F})
is an effective two-particle equation.
Equation (\ref{1F}) is represented diagramatically  in Fig. 1.

 In the model three-particle equation (\ref{1F})
the Green function $ \tau (\sigma _p)$ is designed to maintain
two-particle unitarity. The three dimensional reduction of the Born term
of Eq. (\ref{1F}) will be responsible for maintaining three-particle
unitarity. There exists several unitary reductions of the Born term,
\cite{AAY,Tjon}, here we provide other distinct unitary reductions.

Next, with the three-dimensional Born term,  as in the two-particle case,
we would like to reduce Eq. (\ref{1F}) to a three dimensional form by
performing the integration over the variable $p_0$. During the process
of integration we assume that the $t$ matrix, $T$, and the Born term are
independent of this variable as in the last section. With this
assumption the last term in Eq. (\ref{1F}) is written  as
\begin{eqnarray}
2B \tau T & \equiv & \frac {2i}{(2\pi)^4} \int d \vec p \int _{-\infty}
^\infty  dp_0 \frac {g g  \tau (\sigma _p)
T(\vec p, \vec k,s)}  { (p^2 - m^2 +i0)[(P-q-p)^2-m^2+i0]}, \label{1G1} \\
& = & \frac {2i}{(2\pi)^4} \int d \vec p \int _{-\infty}
^\infty  dp_0 \frac { g g   \tau (\sigma _p)
T(\vec p, \vec k,s)} {(p_0^2-\omega^2_{\vec p}+i0)[(\sqrt s -\omega_{\vec q}
-p_0)^2 -\omega^2_{\vec p+\vec q} +i0]},
\label{1H} \\
& = & \frac {2i}{(2\pi)^4} \int d \vec p \int _{-\infty}
^\infty  dp_0 \frac { g g   \tau (\sigma _p)
T(\vec p, \vec k,s)} {(p_0+\omega_{\vec p} -i0)(p_0-\omega_{\vec p} +i0)}
\nonumber \\
& \times & \frac {1} {(p_0+\omega_{\vec q} -\sqrt s +\omega_{\vec p+\vec q}
-i0)(p_0+\omega_{\vec q} -\sqrt s -\omega_{\vec p+\vec q}
+i0)}. \label{11H}
 \end{eqnarray}

Equation (\ref{1G1}) and the homogeneous term of Eq. (\ref{BS2}) are quite
similar. In both we have two single particle propagators. The difference
is that in Eq. (\ref{BS2}) this propagation occurs in the c. m. system of the
two particles, whereas in Eq. (\ref{1G1}) this propagation occurs in the
c. m. system of the three particles. If we set $(P-q) = (\sqrt s, 0 ,0,0,)$
in Eq. (\ref{1G1}), we arrive at the c. m. frame of two particles and Eqs.
(\ref{1G1}) and (\ref{BS2}) become identical. Because of this, the general
analysis and the final result in both cases are similar. However, the
algebra is more involved in the three-particle case.

The $dp_0$ integral in Eq. (\ref{11H}) is quite similar to that in Eq.
(\ref{E}) and it is evaluated by closing the contour of $p_0$ integration in
the counterclockwise sense along a semicircle in the upper half of the complex
$p_0$ plane at infinity.  The result is given by
\begin{eqnarray}
2B \tau T  & = & -\frac {2}{(2\pi )^3}\int
d\vec p g g   \tau (\sigma _p) T(\vec p, \vec k, s) \nonumber \\
& \times & \biggr[ \frac {1} {(-2\omega_{\vec p})(\omega_{\vec q}-
\omega_{\vec p}-\sqrt s +\omega_{\vec p+\vec q})(\omega_{\vec q}-
\omega_{\vec p}-\sqrt s -\omega_{\vec p+\vec q})} \nonumber \\
& + & \frac {1} {(-2\omega_{\vec p+\vec q})(\omega_{\vec p}-
\omega_{\vec q}+\sqrt s +\omega_{\vec p+\vec q})(-\omega_{\vec q}-
\omega_{\vec p}-\sqrt s -\omega_{\vec p+\vec q}+i0)} \biggr], \\
& = &\frac {2}{(2\pi )^3}\int
\frac{d\vec p}{2\omega _{\vec p}} \frac{g(\omega_{\vec p}+\omega_{\vec p
+\vec q})g}{\omega_{\vec p+\vec q}[(\sqrt s-\omega_{\vec q})^2-
(\omega_{\vec p}+\omega_{\vec p
+\vec q})^2+i0]}  \tau(\sigma_p)T(\vec p, \vec k, s).
\label{1I} \end{eqnarray}
In this equation only the relevant $i0$ which controls the three-particle
unitarity has been shown. The three-dimensional reduction of Eq. (\ref{1F})
is usually written in the form\cite{AAY}
\begin{equation}
T(\vec q, \vec k, s)=2B(\vec q,\vec k,s)+\frac {2}{(2\pi )^3}\int
\frac{d\vec p}{2\omega _p}
B(\vec q, \vec p,s)  \tau (\sigma _p) T(\vec p, \vec k, s).
\label{1K} \end{equation}

Comparing Eq. (\ref{1I}) with the last term on the right-hand-side of
Eq. (\ref{1K}) we identify the Born term of the three-dimensional
scattering integral equation (\ref{1K}) as
\begin{equation}
B(\vec q, \vec p, s) = \frac{g(\omega_{\vec p}+\omega_{\vec p
+\vec q})g}{\omega_{\vec p+\vec q}[(\sqrt s-\omega_{\vec q})^2
-(\omega_{\vec p}+\omega_{\vec p
+\vec q})^2+i0]}.
\label{1L} \end{equation}
Equations (\ref{1K}) and (\ref{1L}) are the desired three-dimensional
instant-form scattering integral equations. These equations follow
from the use of a relativistic three-particle propagator suggested by
Ahmadzadeh and Tjon.\cite{Tjon} However, Eqs. (\ref{1K}) and (\ref{1L}) have
never appeared explicitly in this form before.

Aaron, Amado, and Young\cite{AAY} derived Eq. (\ref{1K}) with the Born term
\begin{equation}
 B(\vec q, \vec p, s) = \frac {g(\omega _{\vec p+\vec q}+\omega _{\vec p}
 +\omega_ {\vec q})g}
{\omega _{\vec p+\vec q}[s-(\omega _{\vec p+\vec q}+\omega _{\vec p}+\omega
_{\vec q})^2+i0]}.
\label{1M} \end{equation}

Both sets - Eqs. (\ref{1K}) and (\ref{1L}), and Eqs. (\ref{1K}) and
(\ref{1M}) -  constitute  linear, three-dimensional, rotationally
invariant and
Lorentz-covariant integral equations for the elastic scattering of one
particle from the bound state (isobar) of the other two.  They satisfy, by
construction, two- and three-particle unitarity. After
a partial-wave projection their solution
can be  easily related to scattering phase-shifts in usual fashion.

The factors $g$'s in Eqs. (\ref{1L}) and
 (\ref{1M}) and in the following are the two-particle potential
form factors as in  Eq. (\ref{W}). The arguments of these form factors
have been suppressed to save space.
We have noted in Sec. II E that the argument should be invariant and be
 the square of the relative momentum in the c. m. frame.
   In the instant form these arguments are explicitly given by\cite{Garc}
\begin{equation}
  {\cal P}^2 = (\omega _{\vec q}+\omega_{\vec p+\vec q})^2/4-p^2/4 -m^2,
  \label{11} \end{equation}
 \begin{equation}
  {\cal Q}^2 = (\omega _{\vec p}+\omega_{\vec p+\vec q})^2/4-q^2/4 -m^2.
\label{12} \end{equation}

The denominator in Eq. (\ref{1M}) has two poles for $\sqrt s = \pm
(\omega _{\vec p+\vec q}+\omega _{\vec p}+\omega_{\vec q}).$ The plus sign
refers to propagation of three particles and is alone responsible for
maintaining the desired unitarity in the three-particle sector. The minus
sign refers to propagation of three antiparticles. Note that the pole
corresponding to the propagation of antiparticles does not materialize in
the physical region: $s>0$.

The present Born term (\ref{1L}) has two poles, too. One of them given by
$\sqrt s =(\omega _{\vec p+\vec q}+\omega _{\vec p}+\omega_{\vec q})$ refers
to the propagation of three particles and should be responsible for maintaining
three-particle unitarity. The other pole is given by $\sqrt s =(\omega_{\vec q}
-\omega _{\vec p+\vec q}-\omega _{\vec p})$  and represents the
propagation of a particle and two antiparticles.   Note that the pole
corresponding to the propagation of antiparticles does not materialize in
the scattering region: $s > 3m$. The requirement of the constraint of
relativistic unitarity for these Born terms is to possess the same residue at
the pole for three-particle propagation at
$\sqrt s =(\omega _{\vec p+\vec q}+\omega _{\vec p}+\omega_{\vec q})$.
This residue of the two Born
terms (\ref{1L}) and (\ref{1M}) at the pole corresponding to the propagation
of three particles are easily shown to be the same $(=gg/2\omega_{\vec q
+\vec p})$.

There is another three-dimensional reduction which has all the
desired features and does not allow antiparticle propagation in the
physical scattering region.  This reduction is given by
\begin{equation}
B(\vec q, \vec p;s)= \frac {g(\omega _{\vec q}+\omega _{\vec p}) g}
{\omega _{\vec q+\vec p}[(\sqrt s - \omega _{\vec q+\vec p})^2
 -(\omega _{\vec q}+\omega _{\vec p})^2+i0]}.
\label{1L1} \end{equation}
This Born term has two poles, too. One of them given by
$\sqrt s =(\omega _{\vec p+\vec q}+\omega _{\vec p}+\omega_{\vec q})$ refers
to the propagation of three particles and is be responsible for maintaining
three-particle unitarity. The other pole is given by $\sqrt s =(
\omega _{\vec p+\vec q}-\omega _{\vec p}-\omega_{\vec q})$  and represents the
propagation of two antiparticles and a particle.   The residue of the two Born
terms (\ref{1M}) and (\ref{1L1}) at the pole corresponding to the propagation
of three particles are easily shown to be the same. Again,
the propagation of the antiparticle does not materialize
in the physical scattering region in the case of the Born term (\ref{1L1}).

In the three-particle Born term we
 could suppress the
antiparticle propagation in the intermediate state. This is achieved by
evaluating the denominator corresponding to antiparticle propagation at the
pole corresponding to particle propagation. Mathematically, this will
constitute in exhibiting the pole for three-particle propagation and the
residue. Consequently,
 the Born terms
(\ref{1L}), (\ref{1M}) and (\ref{1L1}) reduce to the following minimal form
\begin{equation}
B(\vec q, \vec p, s) = \frac{gg}{2\omega _{\vec p+\vec q}
[\sqrt s-\omega _{\vec p}-\omega _{\vec q}-\omega _{\vec p+\vec q} +i0]}.
\label{BA} \end{equation}
This is the simplest Born term which when  used in Eq. (\ref{1K})
will yield unitary result and which does not allow propagation of antiparticles
in the intermediate state.

The Born terms (\ref{1L1})
and (\ref{BA}) are
worth investigating analytically and numerically.

Next, let us consider the non-relativistic limit of these Born terms.
 In this limit the denominator,
$(\sqrt s -\omega _{\vec p+\vec q}-\omega _{\vec p}-\omega_{\vec q})$,
corresponding to three-particle propagation reduces to
$[E- \{(\vec p +\vec q)^2 +\vec p ^2+\vec q ^2\}/2m] $,
where $E$ is the kinetic
energy of the three-particle system in the c. m. frame. In the remaining
factors of the Born term,
$\omega $ reduces to mass $m$ and $\sqrt s$ reduces to $3m$. The arguments
of the form factors $g$'s, given by Eqs. (\ref{11}) and (\ref{12}), reduce
to $(\vec p/2 +\vec q)^2$ and $(\vec q/2 +\vec p)^2$, respectively.
Consequently, all these Born terms reduce to the usual non-relativistic
Amado model\cite{Amado}  Born term given by
\begin{equation}
B(\vec q,\vec p,E) \sim \frac {g[(\vec p/2 +\vec q)^2]g[(\vec q/2 +\vec p)^2]}
{2mE- (\vec p +\vec q)^2 -\vec p ^2-\vec q ^2 +i0}. \label{BBB}
\end{equation}
The phase space of Eq.(\ref{1K}) reduces to the usual
non-relativistic phase space $d\vec p/[2m(2\pi )^3]$ in this limit, and
the present set of equations reduces to the usual Amado model.\cite{Amado}

Aaron, Amado, and Young\cite{AAY} have shown how to calculate breakup
amplitudes from a solution of similar equations. They also have demonstrated
how to include spin
and fermions in the formalism. Similar equations have frequently been used in
pion-nucleon and pion-deuteron scattering problems.\cite{Afnan} Following
their procedure
we could extend these equations to more realistic situations.

\subsection{Light-front equations}

Next we would like to perform the reduction procedure of the last subsection
in the light-front variables.  Specifically, as in Sec. IIC we would like
to perform the $p_-$ integration in Eq. (\ref{1F}) and write a three
dimensional integral equation in terms of variables $(p_+, \vec p_\perp)$
or $(x_p, \vec p_\perp)$. As usual, during the integration we assume that the
$t$ matrix, $T$,  is independent of the variable $p_-$.

For the two particles in the intermediate state in the last term in Fig. 1,
we have
\begin{equation}
p^2-m^2+i0=p_+ \left( p_- -\frac {\vec p_ \perp ^2+m^2-i0}{p_+}\right)
\label{1N} \end{equation}
and
\begin{equation}
(P-q-p)^2 -m^2+i0=(\sqrt s-q_+-p_+)\biggr[ \sqrt s-q_--p_--\frac
{(\vec p_\perp +\vec q_\perp )
^2+m^2-i0}{\sqrt s-q_+-p_+}\biggr] .
\label{1O} \end{equation}

If we use equations (\ref{1N}) and (\ref{1O}),
the last term in Eq.(\ref{1F}) is written explicitly as
\begin{eqnarray}
2B \tau T & \equiv & \frac  {2i}{2(2\pi)^4}\int d \vec p_\perp \int
_{-\infty } ^{\infty } dp_+ \int _{-\infty } ^{\infty }  dp_- \frac
{g  \tau(\sigma _p) g T(p_+,\vec p_\perp ;k_+,\vec k_\perp ;s)}
{p_+(p_- -\frac {\vec p_ \perp ^2+m^2-i0}{p_+})} \nonumber \\
& \times & \frac {1} {(\sqrt s-q_+-p_+)\biggr[ \sqrt s-q_--p_--
\frac{(\vec p_\perp +\vec q_\perp )
^2+m^2-i0}{\sqrt s-q_+-p_+}\biggr] }.
\label{1P} \end{eqnarray}
In Eq. (\ref{1P}), as in Eq. (\ref{J}), the factor of (1/2) before the
integral is the Jacobian of the transformation of integral variables.

The integral in Eq. (\ref{1P}) has the same structure as that in Eq. (\ref{J}),
and the integral over $p_-$ is performed by the technique of contour
integration in a similar fashion in the
complex $p_-$ plane, considering a contour along the real axis from
$-\infty$ to $\infty$ and closing it in the counterclockwise sense along
a semicircle at infinity. After performing this contour integration Eq.
(\ref{1P}) reduces to
\begin{eqnarray}
2B \tau T & = & \frac {2i}{(2\pi)^4}\int d \vec p_\perp \int _0 ^{\sqrt s
-q_+} \frac {dp_+} {2 p_+} \frac {-2\pi i g  \tau(\sigma _p) g
T(p_+,\vec p_\perp ;
k_+,\vec k_\perp ;s)}{(\sqrt s -q_+-p_+)} \nonumber \\
& \times & \frac {1} {\biggr[ \sqrt s -q_--\frac {\vec p_\perp
^2 +m^2}{p_+} -\frac {(\vec p_\perp +\vec q_\perp )^2+m^2}{\sqrt s
-q_+-p_+}+i0\biggr] }.
\label{1Q} \end{eqnarray}
The limits on the integral over $p_+$ in Eq.(\ref{1Q}) appears, as in Eq.
(\ref{L}), as a consequence of including a single pole in the contour.
(See, discussion related to Eq. (\ref{K}) in Sec. IIC.)

If we set the external particle with momentum $q$ on the mass shell, as in
Eq. (\ref{2}), Eq. (\ref{1Q}) can be rewritten as
\begin{eqnarray}
2B \tau T & = & \frac {2}{(2\pi)^3}\int d \vec p_\perp
\int _0 ^{\sqrt s
-q_+} \frac {dp_+}{2p_+} \biggr[\frac {g g }{{(\sqrt s -q_+-p_+)}
[  \sqrt s-   \frac {\vec q_\perp
^2 +m^2}{q_+}  -\frac {\vec p_\perp
^2 +m^2}{p_+} -\frac {(\vec p_\perp +\vec q_\perp )^2+m^2}{\sqrt s
-q_+-p_+}+i0] }\biggr]
 \nonumber \\
& \times &
{  \tau(\sigma _p)  T(p_+,\vec p_\perp ;
k_+,\vec k_\perp ;s)}.
\label{1R} \end{eqnarray}

Equation (\ref{1R})  can also be written in terms of the
momentum fractions, as in the two-particle case, conveniently defined by
\begin{equation}
x_p=p_+/\sqrt s,
\label{1V} \end{equation}
and
\begin{equation}
x_q=q_+/\sqrt s.
\label{1V1} \end{equation}
Using these momentum fractions Eq. (\ref{1R}) can be  rewritten as
\begin{eqnarray}
2B \tau T  & = &   \frac {2}{(2\pi )^3}
 \int _0 ^{1-x_q}  \frac {dx_p}{2x_p} \int d\vec p_\perp
\biggr[  \frac {g g}{(1-x_p-x_q)[  s-
\frac {\vec q_\perp ^2 +m^2}{x_q}  -\frac {\vec p_\perp
^2 +m^2}{x_p} -\frac {(\vec p_\perp +\vec q_\perp )^2+m^2}{1-x_q-x_p}+
i0] } \biggr]
 \nonumber \\ & \times &
\tau (\sigma _p) T(x_p,\vec p_\perp;x_k,\vec k_\perp;s).
\label{1W} \end{eqnarray}
It is tempting to define the quantity in the square bracket of Eq. (\ref{1W})
as the Born term of the following scattering integral equation
\begin{equation}
T = B + B\tau T, \label{2W}
\end{equation}
as in the last subsection. But this identification is not quite to the point
as Eq. (\ref{2W}) with the homogeneous term (\ref{1W}) is of the Volterra
type and not of the usual Fredholm type. A standard Fredholm equation could
be derived by a transformation of variables.

It is convenient to rewrite Eq. (\ref{1W}) in terms of new variable, $x'_p$,
defined by
\begin{equation}
x'_p = \frac {x_p} {1-x_q}.
\label{3W} \end{equation}
In terms of this new variable Eq. (\ref{1W}) is rewritten as
\begin{eqnarray}
2B \tau T  & = &   \frac {2}{(2\pi )^3}
 \int _0 ^{1}  \frac {dx'_p}{2x'_p(1-x'_p)} \int d\vec p_\perp
\biggr[  \frac {g g}{(1-x_q)  (s-
\frac {\vec q_\perp ^2 +m^2}{x_q})  -\frac {\vec p_\perp
^2 +m^2}{x'_p} -\frac {(\vec p_\perp +\vec q_\perp )^2+m^2}{1-x'_p}+
i0 } \biggr]
 \nonumber \\ & \times &
\tau (\sigma _p) T(x_p,\vec p_\perp;x_k,\vec k_\perp;s).
\label{4W} \end{eqnarray}

The quantity $\tau$ of Eq. (\ref{4W}) plays the role of the two-particle
Green function. However, Eq. (\ref{4W})  by itself can not define the Born
term of the scattering integral equation. The three-particle scattering
equation should have the following generic form
\begin{eqnarray}
T(x_q,\vec q_{\perp};x_k,\vec k_{\perp};s) & = & B(x_q,\vec q_{\perp};x_k,
\vec k_{\perp};s)+   \frac {2}{(2\pi )^3} \biggr[
 \int _0 ^{1}  \frac {dx'_p}{2x'_p(1-x'_p)} \int d\vec p_\perp
 \frac {1} {f(x_q,\vec q_{\perp};x'_p,\vec p_{\perp};s)}\biggr]
 \nonumber \\ & \times & B(x_q,\vec q_{\perp};x'_p,\vec p_{\perp};s)
\tau (\sigma _p) T(x'_p,\vec p_\perp;x_k,\vec k_\perp;s),
\label{5W} \end{eqnarray}
with the Born term given by
\begin{equation}
B(x_q,\vec q_{\perp};x'_p,\vec p_{\perp};s)
=\frac {gf(x_q,\vec q_{\perp};x'_p,\vec p_{\perp};s)g}
{(1-x_q)  (s-
\frac {\vec q_\perp ^2 +m^2}{x_q})  +\frac {\vec p_\perp
^2 +m^2}{x'_p} +\frac {(\vec p_\perp +\vec q_\perp )^2+m^2}{1-x'_p}+
i0 },
\label{6W} \end{equation}
where $f(x_q,\vec q_{\perp};x'_p,\vec p_{\perp};s)$ is a function yet to be
determined. This function is cancelled in the homogeneous term of Eq.
(\ref{5W}). This function and hence the Born term can be uniquely determined
by claiming the equivalence between the instant-form three-particle
equations (\ref{1K}) and (\ref{1L}), and light-front three-particle
equations
(\ref{5W}) and (\ref{6W}). Equations (\ref{5W}) and (\ref{6W}) are the
three-particle light-front equations which we  seek.  The homogeneous version
of this equation has recently been solved numerically in a simple
model.\cite{Fred} We also note that Eqs. (\ref{5W}) and (\ref{6W})
are not rotationally invariant in light-front variables.

We shall show by a transformation of variable that the Born term and
the dynamics represented by Eqs. (\ref{5W}) and (\ref{6W})
are essentially the same as those of the instant-form equations
 (\ref{1K}) and (\ref{1L}).

\subsection{Equivalence between light-front and instant-form equations}

In the two-particle problem starting from the BS equation (\ref{BS2}) we
 derived the instant-form and the light-front equations, (\ref{G}) and
 (\ref{N}) by integrating over  the variables
 $p_0$ and $p_-$, respectively. It was demonstrated that these two
forms of equations were equivalent.  We now carry  on the same procedure
in the case of the three-particle system. By claiming the equivalence between
the light-front and instant-form equations we determine the Born term
(\ref{6W}).

In the present case we are really considering the dynamics of two on the
mass shell particles in the three-particle c. m. system. The momentum of
these two particles are given by $(\omega_{\vec p}, \vec p)$ and
$(\omega_{\vec p+\vec q}, -\vec p-\vec q)$. The momentum
fraction of the first particle $x'_p$ is given by
\begin{equation}
x'_p = \frac{\omega_{\vec p}+p_3} {\omega _{\vec p}+\omega _{\vec p+\vec q}
-q_3}.
\label{7W} \end{equation}
We shall see that this transformation, as in the two-particle case, will
relate the three-particle instant-form and light-front equations. The
Jacobian of this transformation is given by
\begin{equation}
\frac{dx'_p}{dp_3}=x'_p(1-x'_p) \frac{\omega _{\vec p}+\omega _{\vec p+
\vec q}} {\omega _{\vec p}\omega _{\vec p+\vec q}}.
\label{8W} \end{equation}
Equations (\ref{7W}) and (\ref{8W}) should be compared with corresponding
equations (\ref{O}) and (\ref{P}) in the two-particle case. They are quite
similar as both represent essentially two-particle dynamics. Recalling
that $x_q$ is defined by Eq. (\ref{1V1}), the Born term (\ref{6W}) is rewritten
as
\begin{equation}
B(x_q,\vec q_{\perp};x'_p,\vec p_{\perp};s)
=\frac {gf(x_q,\vec q_{\perp};x'_p,\vec p_{\perp};s)g}
{(\sqrt s-\omega_{\vec q}-q_3)(\sqrt s-\omega_{\vec q}+q_3)
-\frac {\vec p_\perp
^2 +m^2}{x'_p} -\frac {(\vec p_\perp +\vec q_\perp )^2+m^2}{1-x'_p}+
i0 }.
\label{9W} \end{equation}
If we use transformation (\ref{7W}) it is straightforward to show that
\begin{equation}
\frac {\vec p_\perp
^2 +m^2}{x'_p} +\frac {(\vec p_\perp +\vec q_\perp )^2+m^2}{1-x'_p}
=(\omega_{\vec p}+\omega_{\vec p +\vec q})^2 - q_3^2.
\label{10W} \end{equation}
Provided that under the transformation (\ref{7W}) the function
$f(x_q,\vec q_{\perp};x'_p,\vec p_{\perp};s)$ transforms according to
\begin{equation}
f(x_q,\vec q_{\perp};x'_p,\vec p_{\perp};s) \to \frac
{\omega_{\vec p}+\omega_{\vec p +\vec q}} {\omega_{\vec p +\vec q}},
\label{11W} \end{equation}
with Eq. (\ref{10W}) the Born term of Eq. (\ref{9W})
reduces to
\begin{equation}
B(x_q,\vec q_{\perp};x'_p,\vec p_{\perp};s) \to \frac {g(\omega_{\vec p}+
\omega_{\vec p +\vec q})g}
{\omega_{\vec p +\vec q}[(\sqrt s -\omega_{\vec q})^2
-(\omega_{\vec p}+\omega_{\vec p +\vec q})^2+i0]},
\label{12W} \end{equation}
which is identical to the Born term (\ref{1L}).
We note that there is not only a
one-to-one correspondence between the Born terms of Eqs. (\ref{1L}) and
(\ref{6W}) under transformation (\ref{7W}), but because of Eqs. (\ref{8W})
and (\ref{11W})
the phase spaces of these two equations are also identical, explicitly
\begin{equation}
\frac {dx'_p}{2x'_p(1-x'_p)}d\vec p_{\perp}
\frac {1} { f(x_q,\vec q_{\perp};x'_p,\vec p_{\perp};s)} \to
\frac{d\vec p}{2\omega_{\vec p}}.
\end{equation}
This proves the complete equivalence between the three-particle
light-front and
instant-form dynamical equations, provided that the function
$f(x_q,\vec q_{\perp};x'_p,\vec p_{\perp};s)$ obeys Eq.
(\ref{11W}) under this transformation. Hence one can perform a
numerical solution of the light-front dynamical equations (\ref{5W})
and (\ref{6W}) via the equivalent instant-form equations (\ref{1L})
and (\ref{1K}). The instant-form equations (\ref{1K}) and (\ref{1L})
are rotationally invariant. The present demonstration of
equivalence guarantees that, as in the two-particle case, the
light-front equations (\ref{5W}) and (\ref{6W}), though not
manifestly rotationally invariant, will yield unitary and
rotationally invariant results in usual instant-form variables.
In short, we have derived the sets of equations given
equivalently, by Eqs. (\ref{1K}) and
(\ref{1L}), and by Eqs. (\ref{5W}) and (\ref{6W}), in the instant-form
and light-front formalisms, respectively.

However, we could not find an explicit expression for the function
$f(x_q,\vec q_{\perp};x'_p,\vec p_{\perp};s)$. We have Eq. (\ref{11W})
subject to the transformation (\ref{7W}).

Fuda\cite{Fuda} also derived three-particle light-front
equations for spinless meson-baryon scattering using  two-particle
isobars.  The homogeneous three-particle equation derived by him is
identical to the homogeneous version of present Eqs. (\ref{5W}) and
(\ref{6W}), though his Born-term is different from the
present work in containing  $\Theta $-functions and using $
f(x_q,\vec q_{\perp};x'_p,\vec p_{\perp};s) =1$. The present light-front
equations (\ref{5W}) and (\ref{6W}) are also distinct from the equations
of Bakker et al.\cite{Bakk} To the best of our knowledge the present
light-front equations (\ref{5W}) and (\ref{6W}) are unique in being related
to a set of instant-form three-particle equations.

\section{SUMMARY}

We have provided general prescriptions for reducing four-dimensional BS like
scattering integral equations to three-dimensional scattering integral
equations of the LS\cite{Lipp} form. The resultant equations,
by construction, are linear,  relativistically covariant, possess correct
nonrelativistic limit and yield unitary result.
We have applied the present procedure to the two- and the three-particle
systems. In the three-particle system the dynamical equations
are written using analogy to the nonrelativistic Amado model\cite{Amado}
and treating the particles to be spinless bosons.

We have carried out the present  procedure in two different sets of
variables -
the instant-form and light-front variables. Our procedure is to carry on the
integration over the time component of momentum in the intermediate state phase
space while assuming, as usual, that the integrand is independent of the time
component of momentum. In the instant form this implies an integration over
$p_0$, and in the light-front formalism this implies an integration over
$p_-$.  As the starting four-dimensional integral equations satisfy constraints
of relativistic unitarity and covariance, the reduced equations also do so as
there are no approximations which could destroy these virtues.

In the case of the two-particle system the present procedure yields the
well known light-front equation first derived by Weinberg\cite{Wein} and
instant-form equation first derived by Blankenbecler and Sugar\cite
{BlS}. These two types of equations represent the same dynamics and are
demonstrated to be related by a transformation of variables and hence are
equivalent to each other. Consequently, the light-front
two-particle equation, though not rotationally invariant in the light-front
variables, will yield rotationally invariant result when expressed in
instant-form variables.

The equivalence between the two types of two-particle equations allow one
to perform practical partial-wave calculations using the light-front
equation (\ref{N}). We have shown in Sec. IID how to construct a
light-front potential $V(x_q,\vec q;x_p,\vec p;s)$ from a spherically
symmetric partial-wave potential $V_l (q,p,s)$ using Eq. (\ref{V}).
Once this potential is constructed the three-dimensional light-front
equation  (\ref{N}) could be numerically solved to yield the same
result as obtained by a direct solution of the one-dimensional partial
wave instant-form equation.  As the light-front equation continues to
be three-dimensional, its solution may imply added numerical complications.
But the final result, when expressed in terms of the physical instant-form
variables, is identical to that obtained by the direct solution of the
one-dimensional partial-wave instant-form equation. This is demonstrated
explicitly in an analytic separable potential model. Following this procedure
one can circumvent
the difficulties with the angular momentum projection of the light-front
dynamical equations in carrying out a numerical calculation.

In the case of the three-particle system the present procedure yields new
practical three-dimensional
scattering equations using the instant-form and light-front variables.
We derive several instant-form equations
satisfying constraints of relativistic unitarity and  covariance. Of these,
the sets of Eqs.  (\ref{1K}) and (\ref{BA}), Eqs. (\ref{1K}) and (\ref{1L1}),
and Eqs.  (\ref{1K}) and (\ref{1L}) are new and deserve more attention in the
future.
Of these sets Eqs.  (\ref{1K}) and (\ref{1L}) follow as a consequence of the
use  of a relativistic propagator suggested long ago\cite{Tjon}. However,
Eqs.  (\ref{1K}) and (\ref{1L}) in this form have never appeared in the
literature before.  The present set of light-front
three-particle equations (\ref{5W}) and (\ref{6W}) are shown to be related
by a transformation of variables to the instant-form equations (\ref{1K})
and (\ref{1L}), and hence the two sets represent identical dynamics. To the
best of our knowledge this is the first time that a relation is established
via a transformation of variables between three-particle equations in
instant-form and light-front variables. The solutions of these
two sets of equations are supposed to be identical as in the two-particle
case.  Distinct relativistic, unitary three-particle instant-form equations
are provided by other authors.\cite{AAY}
 The three-particle light-front equations (\ref{5W}) and (\ref{6W}),
are, however, distinct from the three-particle light-front equations of
Refs. \cite{Fuda,Bakk}.

The present set of instant-form three-particle
equations can easily be used for studying relativistic effects on
both bound state and scattering of the three-nucleon system, or
studying intermediate energy pion-deuteron or pion-nucleon scattering.
In the following paper we present a numerical application of the present set
of equations to the study of the relativistic effect in the trinucleon
system.\cite{our}

The work is supported in part by the Conselho Nacional de
Desenvolvimento - Cient\'\i fico e Tecnol\'ogico (CNPq) of Brasil.


\newpage

\noindent {\bf Figure Caption}

1. The diagramatic form of the three-particle scattering equation (\ref{1F}).
Each single-line represents propagation of a particle, each double-line
represents propagation of the bound state or isobar of two particles.

\unitlength=1.00mm
\special{em:linewidth 0.4pt}
\linethickness{0.4pt}
\begin{picture}(157.00,135.00)
\put(40.00,85.00){\circle{14.00}}
\put(25.00,92.00){\line(1,0){31.00}}
\put(25.00,78.00){\line(1,0){31.00}}
\put(65.00,92.00){\line(1,0){26.00}}
\put(65.00,78.00){\line(1,0){26.00}}
\put(100.00,92.00){\line(1,0){56.00}}
\put(100.00,78.00){\line(1,0){56.00}}
\put(140.00,85.00){\circle{14.00}}
\put(65.00,90.00){\line(1,0){10.00}}
\put(75.00,90.00){\line(3,-5){6.00}}
\put(81.00,80.00){\line(1,0){10.00}}
\put(100.00,90.00){\line(1,0){10.00}}
\put(110.00,90.00){\line(3,-5){6.00}}
\put(116.00,80.00){\line(1,0){19.00}}
\put(145.00,80.00){\line(1,0){11.00}}
\put(30.00,96.00){\makebox(0,0)[ct]{(P-q)}}
\put(50.00,96.00){\makebox(0,0)[ct]{k}}
\put(69.00,96.00){\makebox(0,0)[ct]{(P-q)}}
\put(86.00,96.00){\makebox(0,0)[ct]{k}}
\put(86.00,75.00){\makebox(0,0)[cb]{(P-k)}}
\put(69.00,75.00){\makebox(0,0)[cb]{q}}
\put(30.00,75.00){\makebox(0,0)[cb]{q}}
\put(50.00,75.00){\makebox(0,0)[cb]{(P-k)}}
\put(104.00,96.00){\makebox(0,0)[ct]{(P-q)}}
\put(104.00,75.00){\makebox(0,0)[cb]{q}}
\put(125.00,96.00){\makebox(0,0)[ct]{p}}
\put(125.00,75.00){\makebox(0,0)[cb]{(P-p)}}
\put(150.00,75.00){\makebox(0,0)[cb]{(P-k)}}
\put(149.00,96.00){\makebox(0,0)[ct]{k}}
\put(107.00,84.00){\makebox(0,0)[cc]{(P-p-q)}}
\put(60.00,85.00){\makebox(0,0)[cc]{$=$}}
\put(95.00,85.00){\makebox(0,0)[cc]{$+$}}
\put(45.00,80.00){\line(1,0){11.00}}
\put(25.00,90.00){\line(1,0){10.00}}
\end{picture}

\end{document}